\documentclass[twocolumn,showpacs,preprintnumbers,amsmath,amssymb]{./revtex4-1}

\usepackage{makeidx}     
\usepackage{graphicx}    
\usepackage{dcolumn}
\usepackage{bm}
\usepackage{color}
\usepackage{amssymb,amsfonts,amsmath}
\usepackage{equations}
\usepackage{url}
\usepackage{framed}
\usepackage{amsmath,amssymb,amscd,ulem}
\usepackage{bm}
\usepackage[hypertex]{hyperref}
\usepackage{comment}

\newcommand{\ProbP}{\mathbb{P}}
\newcommand{\ProbQ}{\mathbb{Q}}

\newcommand{\ProbT}{\mathbb{T}}

\newcommand{\LF}{\mathbb{H}}
\newcommand{\defeq}{:=}
\newcommand{\CMF}{\Psi}
\newcommand{\LTGR}{\psi}
\newcommand{\MI}{\mathcal{I}}
\newcommand{\average}[1]{\left<#1 \right>}
\newcommand{\KL}[2]{\mathcal{D}[#1||#2]}
 \DeclareMathAlphabet{\mathpzc}{OT1}{pzc}{m}{it}
 \newcommand{\xpzc}{\mathpzc{X}}
\newcommand{\ypzc}{\mathpzc{Y}}
\newcommand{\zpzc}{\mathpzc{Z}}

\newcommand{\pop}{\mathcal{N}}

\newcommand{\eqnref}[1]{eq. (\ref{#1})}

\newcommand{\fgref}[1]{Fig. \ref{#1}}
\newcommand{\submit}[1]{}

\newcounter{mykaunta}
\newcommand{\thekaunta}{\arabic{mykaunta}}
\newcommand{\mykaunta}{\refstepcounter{mykaunta}%
           \thekaunta}

\begin{document}


\preprint{APS/123-QED}

\title{Fluctuation Relations of Fitness and Information in Population Dynamics}

\author{Tetsuya J. Kobayashi}
\author{Yuki Sughiyama}
 \email{tetsuya@mail.crmind.net}
 \homepage{http://research.crmind.net/}
\affiliation{%
${}^{1}$:Institute of Industrial Science, the University of Tokyo, 4-6-1 Komaba Meguro-ku, Tokyo 153-8505, Japan.
}%

\date{\today}

\pacs{Valid PACS appear here}

\begin{abstract}
Phenotype-switching with and without sensing environment is a ubiquitous strategy of organisms to survive in fluctuating environment. 
Fitness of a population of organisms with phenotype-switching may be constrained and restricted by hidden relations as the entropy production in a thermal system with and without sensing and feedback is well-characterized via fluctuation relations (FRs) . 
In this work, we derive such FRs of fitness together with an underlying information-theoretic structure in selection.
By using path-integral formulation of a multi-phenotype population dynamics, we clarify that the optimal switching strategy is characterized as a consistency condition for time-forward and backward path probabilities. 
Within the formulation, the  selection is regarded as passive information compression, and the loss of fitness from the optimal strategy is shown to satisfy various FRs that constrain the average and fluctuation of the loss.
These results are naturally extended to the situation that organisms can use an environmental signal by actively sensing the environment.
FRs of fitness gain by sensing are derived in which the multivariate mutual information among the phenotype, the environment and the signal plays the role to quantify the relevant information in the signal for fitness gain.\\
\textbf{Submitted to PRL on 25/Jul/2014; resubmitted to PRL for revision on 10/Apr/2015.}
\end{abstract}
\maketitle

\submit{\section{Introduction}}
Phenotype-switching is a strategy of living systems to survive in stochastically changing environment\cite{Levins:1968tc}. 
Even if no environmental information is available,  diversification of phenotypes by stochastic switching (known as bethedging) can lead to gain of fitness when a subpopulation with a resistant phenotype can survive in a harsh environment to feed the next generation\cite{Haccou:1995tf,Kussell:2005dg,deJong:2011iv}.
If environmental signal that conveys information of the environment is exploitable, further gain of fitness is possible by switching into the phenotypes adapted to the future environment (known as decision-making)\cite{Perkins:2009cg,Kobayashi:2012ji,Rivoire:2011fy}.
Ubiquitous observations of phenotype-switching and environmental sensing in living systems from higher organisms down to bacteria implies its actual fitness advantage over the diversification loss and the metabolic load of switching and sensing mechanisms\cite{Balaban:2004bq,Jayaraman:2008vo,Wakamoto:2013hb,BenJacob:2010ii}.

The fitness gain enjoyed by switching and sensing, however, must be constrained by the environmental statistics and the sensed information.
On the one hand, previous investigations clarified such constrains for average fitness gain at least in specific situations\cite{Haccou:1995tf,Bergstrom:2004um,Kussell:2005dg,DonaldsonMatasci:2010ie,Rivoire:2011fy,Frank:2012kq,Rivoire:2014kt}.
On the other hand, the similarity between evolutionary dynamics and statistical physics\cite{Iwasa:1988ux,deVladar:2011kz,Qian:2014ha} suggests that more general relations may exist as the series of fluctuation relations (FRs) characterize not only the average but also the fluctuation of entropy production in a thermal system with and without sensing and feedback\cite{Seifert:2012es, Sagawa:2012wi}.
Finding such relations is crucial to understand the constraints and predicability of adaptive dynamics of organisms in ever changing environment (fitness seascapes)\cite{Mustonen:2010ig}.
In this work, by using a path-wise (path-integral) formulation of the dynamics of growing population with multi-phenotypes\cite{Leibler:2010jx,Wakamoto:2012hx,Bianconi:2012kz,Oizumi:2013ba}, we reveal such relations of fitness together with the underlying information-theoretic structure of selection.

\begin{figure}
\includegraphics[width=\linewidth]{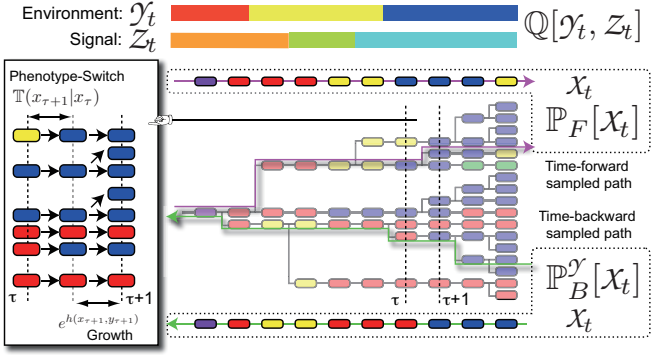}
\caption{A schematic diagram of the population dynamics with phenotype switching\cite{Leibler:2010jx}.  Note that the actual model accounts for sufficiently large population, and the lineage of cells in the figure illustrates behavior of only a small subset of the population.}
\label{fig1}
\end{figure}

\submit{\section{Basic Formulation}}
Let $x_{t} \in \mathcal{S}_{x}$ and $y_{t} \in \mathcal{S}_{y}$ be a phenotype of a living organism and a state of environment at time $t$, respectively.
For simplicity, possible phenotypic and environmental states are assume to be discrete. 
We also define paths (histories) of phenotype and environment up to time $t$ as
$\xpzc_{t} \defeq \{x_{\tau}|\tau \in [0, t]\}$ and $\ypzc_{t} \defeq \{y_{\tau}|\tau \in [0, t]\}$, respectively.
The population size of organisms in a phenotype $x$ at time $t$ under a realization of an environmental path $\ypzc_{t}$ is denoted as $\pop_{t}^{\ypzc}(x)$. 
When the population size is sufficiently large for all $x$, $\pop_{t}^{\ypzc}(x)$ can be approximated to be continuous as in \cite{Haccou:1995tf,Leibler:2010jx}.
Phenotype of an organism, in general, switches stochastically over time depending on its state.
The switching dynamics is modeled, for example, by a Markov transition probability $\ProbT$ where $\sum_{x'}\ProbT(x'|x)=1$ as in \cite{Leibler:2010jx}. 
In addition, an organism with a phenotype $x$ under an environmental state $y$ is assumed to duplicate asexually to produce its $e^{h(x,y)}-1$ copies on average within the unit time interval   where $h: \mathcal{S}_{x} \times \mathcal{S}_{y} \to \mathbb{R}$. 
Then, the time-discrete dynamics of the population size, $\pop_{t}^{\ypzc}(x)$, can be described (\fgref{fig1}) as
\begin{align}
\pop_{t+1}^{\ypzc}(x')=e^{h(x',y_{t+1})}\sum_{x}\ProbT(x'|x)\pop_{t}^{\ypzc}(x). \label{eq:popdyn}
\end{align}
Cumulative fitness of the population at $t$ under an environmental path $\ypzc_{t}$ is defined as
$\CMF[\ypzc_{t}] := \ln\frac{\sum_{x'}\pop_{t}^{\ypzc}(x')}{\sum_{x}\pop_{0}(x)}$.
If the environmental path follows a path probability $\ProbQ[\ypzc_{t}]$, we can define the environmental ensemble average of the cumulative fitness as $\average{\CMF_{t}} \defeq \average{\CMF[\ypzc_{t}]}_{\ProbQ[\ypzc_{t}]}$. 
Moreover, with additional assumptions on $\ypzc_{t}$ and $\ProbT$, the time-average of the cumulative fitness, $\LTGR(t)\defeq \average{\CMF_{t}}/t$ can also reflect  temporal averaging of long-term growth under one realization of the environment as $\lim_{t \to \infty} \frac{1}{t}\CMF[\ypzc_{t}] = \lim_{t \to \infty}\LTGR(t)$\cite{Haccou:1995tf}. 
In this work, finite $t$ is considered.

All formulations and definitions can be naturally extended for the situation where an environmental signal $z_{t} \in \mathcal{S}_{z}$ is available. 
Let $\zpzc_{t}\defeq \{z(\tau)|\tau\in[0,t]\}$ be the path of the signal, and $\ProbQ[\ypzc_{t},\zpzc_{t}]$ be the joint paht-probability of the environment and the signal. 
The population dynamics with the signal can be obtained by simply replacing $\ProbT(x'|x)$ with $\ProbT(x'|x,z')$   as 
\begin{align}
\pop_{t+1}^{\ypzc,\zpzc}(x')=e^{h(x', y_{t+1})}\sum_{x}\ProbT(x'|x,z_{t+1})\pop_{t}^{\ypzc,\zpzc}(x).\label{eq:popdyn2}
\end{align}
We also define the cumulative fitness and its average as $\CMF[\ypzc_{t},\zpzc_{t}] := \ln\frac{\sum_{x'}\pop_{t}^{\ypzc,\zpzc}(x')}{\sum_{x}\pop_{0}(x)}$, and $\average{\CMF_{t}} \defeq \average{\CMF[\ypzc_{t},\zpzc_{t}]}_{\ProbQ[\ypzc_{t},\zpzc_{t}]}$.

\submit{\section{Path integral formulation}}
As indicated in \cite{Leibler:2010jx,Wakamoto:2012hx, Sughiyama:2015cf}, the total population size at time $t$, $\pop_{t}^{\ypzc}\defeq \sum_{x'}\pop_{t}^{\ypzc}(x')$, can be described with a path integral formulation. Let define a time-forward path probability of phenotype without sensing as $\ProbP_{F}[\xpzc_{t}] \defeq \prod_{\tau=0}^{t-1}\ProbT(x_{\tau+1}|x_{\tau})\ProbP_{F}(x_{0})$ where $\ProbP_{F}(x_{0}) \defeq \pop_{0}^{\ypzc}(x_{0})/\pop_{0}^{\ypzc}$.
In addition, a path-wise (historical) fitness of a phenotypic path $\xpzc_{t}$ under an environmental path $\ypzc_{t}$ is defined as $\LF[\xpzc_{t},\ypzc_{t}]\defeq \sum_{\tau=0}^{t-1}h(x_{\tau+1}, y_{\tau+1})$. 
Then, the population size of organisms that experience the phenotypic path $\xpzc_{t}$ under $\ypzc_{t}$ at time $t$ is 
$\pop[\xpzc_{t},\ypzc_{t}] =e^{\LF[\xpzc_{t},\ypzc_{t}]}\ProbP_{F}[\xpzc_{t}]\pop_{0}$\cite{SM:2014,Sughiyama:2015cf}.
Thus, we have $\pop^{\ypzc}_{t}=\sum_{\xpzc_{t}}\pop[\xpzc_{t},\ypzc_{t}]$ and 
\begin{align}
\CMF[\ypzc_{t}]= \ln\frac{\pop_{t}^{\ypzc}}{\pop_{0}^{\ypzc}}=\ln\average{e^{\LF[\xpzc_{t},\ypzc_{t}]}}_{\ProbP_{F}[\xpzc_{t}]}.\label{eq:CMFF}
\end{align}
Because of this representation, $\CMF[\ypzc_{t}]$ can also be represented variationally as 
\begin{align}
\CMF[\ypzc_{t}] = \max_{\ProbP[\xpzc_{t}]}\left[\average{\LF[\xpzc_{t},\ypzc_{t}]}_{\ProbP}-\KL{\ProbP}{\ProbP_{F}} \right], \label{eq:CMFvar}
\end{align}
where $\KL{\ProbP}{\ProbP_{F}} \defeq \sum_{\xpzc_{t}}\ProbP[\xpzc_{t}]\ln \frac{\ProbP[\xpzc_{t}]}{\ProbP_{F}[\xpzc_{t}]}$ is the Kullback-Leibler divergence (KLD) between 
$\ProbP$ and $\ProbP_{F}$\cite{SM:2014,Kullback:1951va,Sughiyama:2015cf}.
This variational problem can be attained by a retrospective (time-backward) path probability \cite{Baake:2006ek,Sughiyama:2015cf} defined as
\begin{align}
\ProbP_{B}^{\ypzc}[\xpzc_{t}] &= e^{\LF[\xpzc_{t},\ypzc_{t}]-\CMF[\ypzc_{t}]}\ProbP_{F}[\xpzc_{t}]. \label{eq:backward}
\end{align}
With this backward path probability, we have 
\begin{align}
\CMF[\ypzc_{t}] =\average{\LF[\xpzc_{t},\ypzc_{t}]}_{\ProbP_{B}^{\ypzc}}-\KL{\ProbP_{B}^{\ypzc}}{\ProbP_{F}} \label{eq:CMFB}.
\end{align}
This path-wise formulation generally holds for more general $\ProbP_{F}$ that may not be generated by Markov processes. 
In the following, therefore, we consider that $\ProbP_{F}[\xpzc_{t}]$ can be any path probabilities over the phenotypic history, $\xpzc_{t}$.

The forward and backward path probabilities, $\ProbP_{F}[\xpzc_{t}]$ and $\ProbP_{B}^{\ypzc}[\xpzc_{t}]$, have an obvious interpretation. 
$\ProbP_{F}[\xpzc_{t}]$ is the probability to observe an organism with a phenotypic history $\xpzc_{t}$ when we randomly sample an organism from the initial population at $t=0$ and track it in a time-forward manner (\fgref{fig1}). 
When the tracked organism duplicates, we choose one of the two daughters randomly. 
$\ProbP_{B}[\xpzc_{t}]$, in contrast,  is the probability to observe $\xpzc_{t}$ when we randomly sample an organism from the final population at $t$ and track it back retrospectively (\fgref{fig1}). 
Because the backward path probability is defined for a fixed environmental history, $\ypzc_{t}$, we can also define a joint path probability as $\ProbP_{B}^{J}[\xpzc_{t},\ypzc_{t}] \defeq \ProbP_{B}^{\ypzc}[\xpzc_{t}]\ProbQ[\ypzc_{t}]$. 
We can also obtain a marginal path probabilities as $\ProbP_{B}^{M}[\xpzc_{t}]\defeq \sum_{\ypzc_{t}} \ProbP_{B}^{J}[\xpzc_{t},\ypzc_{t}]$.

\begin{figure}
\includegraphics[width=\linewidth]{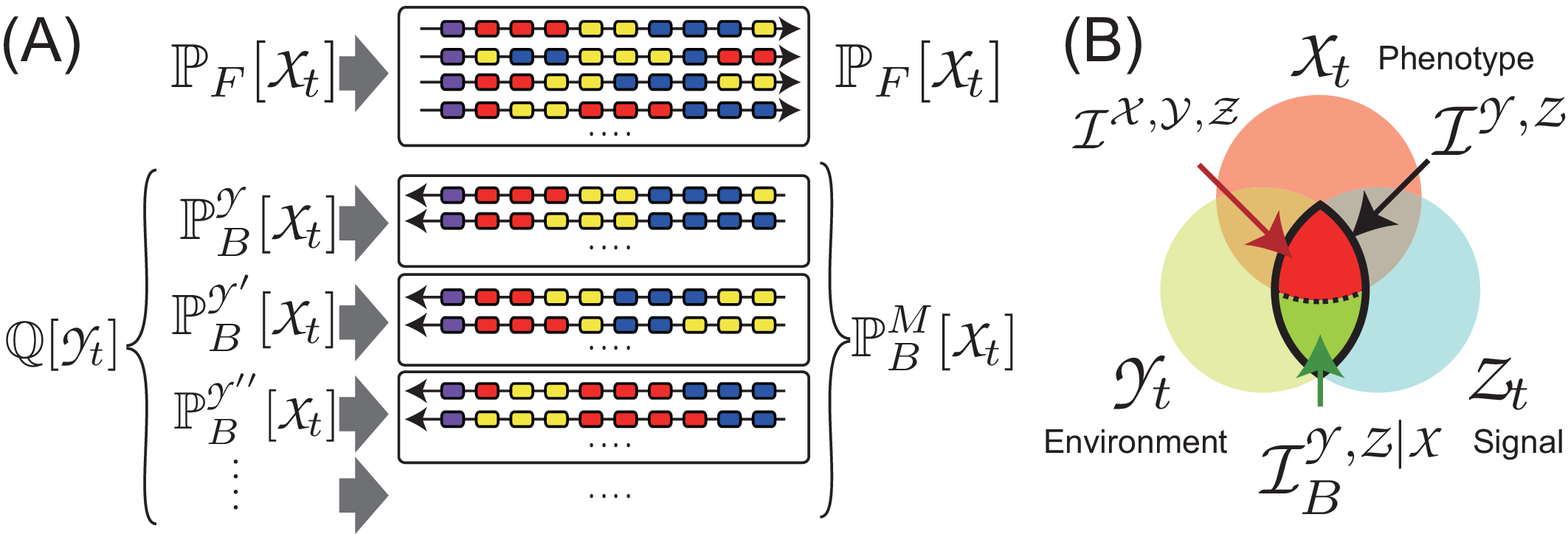}
\caption{(A) A schematic diagram of the forward, backward and the marginal path-ensembles, $\ProbP_{F}[\xpzc_{t}]$,  $\ProbP_{B}^{\ypzc}[\xpzc_{t}]$, and $\ProbP_{B}^{M}[\xpzc_{t}]$. 
(B) Venn's diagram for information among phenotype, environment, and signal. 
The region circled in bold black is $\MI^{\ypzc,\zpzc}$, and those in green and red foreground are $\breve{\MI}_{B}^{\ypzc,\zpzc|\xpzc}$ and $\breve{\MI}_{B}^{\xpzc,\ypzc,\zpzc}$, respectively.}
\label{fig2}
\end{figure}

\submit{\section{Information-theoretic structure in growing population}}
With these probabilities, we have the average of the cumulative fitness as
\begin{align}
\average{\CMF_{t}}  & =\average{\LF[\xpzc_{t},\ypzc_{t}]}_{\ProbP_{B}^{J}}-\MI_{B}^{\xpzc,\ypzc}-\KL{\ProbP_{B}^{M}}{\ProbP_{F}},\label{eq:avCMF}
\end{align}
where $\MI_{B}^{\xpzc,\ypzc}$ is a backward mutual information between $\xpzc_{t}$ and $\ypzc_{t}$ defined as $\MI_{B}^{\xpzc,\ypzc} \defeq \sum_{\xpzc_{t},\ypzc_{t}}\ProbP_{B}^{J}[\xpzc_{t},\ypzc_{t}]\ln \frac{\ProbP_{B}^{J}[\xpzc_{t},\ypzc_{t}]}{\ProbP_{B}^{M}[\xpzc_{t}]\ProbQ[\ypzc_{t}]}$\cite{SM:2014}.
This relation generally holds for any strategy of phenotypic switching.
Among them, we focus on a special strategy, $\hat{\ProbP}_{F}[\xpzc_{t}]$, that maximizes $\average{\CMF_{t}}$. 
Because $\average{\CMF_{t}}$ is concave with respect to $\ProbP_{F}$, $\hat{\ProbP}_{F}[\xpzc_{t}]$ is unique in the convex space of path probabilities. 
However $\hat{\ProbP}_{F}[\xpzc_{t}]$ may not exists within biologically realistic class of path probability, e.g., ones generated by Markov or causal processes.
If exits, this strategy can be regarded as the strategy that have adapted evolutionary to the environment defined by $\ProbQ[\ypzc_{t}]$.
Even if not, $\hat{\ProbP}_{F}[\xpzc_{t}]$ and the corresponding $\average{\hat{\CMF}}\defeq \max_{\ProbP_{F}}\average{\CMF}$ still plays important roles as the bound of fitness in the FRs derived in the following.
Such optimal strategy must satisfy the stationary condition, $\delta \average{\hat{\CMF}_{t}}=0$, for any perturbation of the strategy $\delta \ProbP_{F}$ around $\hat{\ProbP}_{F}$. 
The condition can be explicitly represented as 
$\delta \average{\hat{\CMF}_{t}} = \average{\sum_{\ypzc_{t}} \frac{\hat{\ProbP}_{B}^{\ypzc}[\xpzc_{t}]\ProbQ[\ypzc_{t}]}{\hat{\ProbP}_{F}[\xpzc_{t}]}}_{\delta \ProbP_{F}}= \average{\frac{\hat{\ProbP}_{B}^{M}[\xpzc_{t}]}{\hat{\ProbP}_{F}[\xpzc_{t}]}}_{\delta \ProbP_{F}}=0$\cite{SM:2014}. 
From this equation, we obtain a consistency condition between the forward and backward probabilities  as 
\begin{equation}
{\hat{\ProbP}^{M}_{B}}[\xpzc_{t}]=\sum_{\ypzc_{t}} \hat{\ProbP}_{B}^{\ypzc}[\xpzc_{t}]\ProbQ[\ypzc_{t}]=\hat{\ProbP}_{F}[\xpzc_{t}]. \label{eq:consistency}
\end{equation}
The consistency condition requires no time-directionality in the phenotypic paths in the following sense.
When we sample phenotypic paths in the time-forward manner, we have the ensemble of paths, $\ProbP_{F}[\xpzc_{t}]$, that contains no influence from the environment. 
When we sample paths in the time-backward manner without observing the environment, $\ypzc_{t}$, we have another ensemble of paths that follows the marginal backward path probability $\hat{\ProbP}_{B}^{M}[\xpzc_{t}]$(\fgref{fig2} (A)).
While the forward and the backward path probabilities are the same marginally under the consistency condition, the selection induces correlation between the backward phenotypic dynamics and the environmental history.
This fact is quantitatively described by the optimal cumulative fitness obtained from \eqnref{eq:avCMF} as 
\begin{align}
\average{\hat{\CMF}_{t}} = \average{\LF[\xpzc_{t},\ypzc_{t}]}_{{\hat{\ProbP}^{J}_{B}}} -\hat{\MI}_{B}^{\xpzc,\ypzc},\label{eq:maxavCMF}
\end{align}
where $\hat{\MI}_{B}^{\xpzc,\ypzc}$ measures the correlation.
This form of the optimal cumulative fitness can be further represented as another type of  variational problem\cite{SM:2014} as
\begin{align}
 \average{\hat{\CMF}_{t}}  = \max_{\ProbP^{\ypzc}_{B}[\xpzc_{t}]}\left[\average{\LF[\xpzc_{t},\ypzc_{t}]}_{{\ProbP^{\ypzc}_{B}[\xpzc_{t}]\ProbQ[\ypzc_{t}]}} -  \MI^{\xpzc,\ypzc}_{B}\right].\label{eq:maxavCMFvar}
\end{align}
From the information-theoretic viewpoint, this is equivalent to the lossy soft compression or encoding of the environmental history, $\ypzc_{t}$, into the phenotypic history, $\xpzc_{t}$, under a utility measure $\LF$ (or equivalently distortion measure  $-\LF$)\cite{Cover:2012ub}.
The environmental history is composed of information relevant to and nothing to do with increase of the path-wise fitness $\LF$.
The variational form of $\average{\hat{\CMF}_{t}}$ indicates that only information relevant for increasing $\LF$ is imprinted or encoded into the phenotype history, and the optimal  $\hat{\ProbP}^{\ypzc}_{B}[\xpzc_{t}]$ is regarded as the optimal encoder.
This relation clarifies that selection can be regarded as a kind of passive information processing, and the backward mutual information, $\hat{\MI}^{\xpzc,\ypzc}_{B}$,  quantifies the information encoded by the selection.

\submit{\section{Fluctuation Relations}}
From the consistency condition (\eqnref{eq:consistency}), we have $\sum_{\ypzc_{t}}\hat{\ProbP}_{B}^{\ypzc}[\xpzc_{t}]\ProbQ[\ypzc_{t}]/\hat{\ProbP}_{F}[\xpzc_{t}]=\sum_{\ypzc_{t}}e^{\LF[\xpzc_{t},\ypzc_{t}]-\hat{\CMF}[\ypzc_{t}]}\ProbQ[\ypzc_{t}]=1$. This implies that $\hat{\ProbP}_{B}^{\xpzc}[\ypzc_{t}]\defeq \hat{\ProbP}_{B}^{J}[\xpzc_{t},\ypzc_{t}]/\hat{\ProbP}_{B}^{M}[\xpzc_{t}]= e^{\LF[\xpzc_{t},\ypzc_{t}]-\hat{\CMF}[\ypzc_{t}]}\ProbQ[\ypzc_{t}]$ holds for $\xpzc_{t} \in \mathrm{Supp}[\hat{\ProbP}_{F}]\defeq \{\xpzc_{t}|\hat{\ProbP}_{F}[\xpzc_{t}]\neq 0\}$. 
By taking average with any $\ProbP_{F}[\xpzc_{t}]$ sharing the same support with $\hat{\ProbP}_{F}$, we can easily see that the fitness loss of a suboptimal strategy defined as $\Delta \CMF[\ypzc_{t}] \defeq \hat{\CMF}[\ypzc_{t}]-\CMF[\ypzc_{t}]$ satisfies the following detailed FR:
\begin{align}
e^{-\Delta \CMF[\ypzc_{t}]} =\frac{\hat{\ProbP}_{B}^{\ypzc}[\xpzc_{t}]}{\hat{\ProbP}_{F}[\xpzc_{t}]}\frac{\ProbP_{F}[\xpzc_{t}]}{\ProbP_{B}^{\ypzc}[\xpzc_{t}]} =\frac{\average{\hat{\ProbP}_{B}^{\xpzc}[\ypzc_{t}]}_{\ProbP_{F}[\xpzc_{t}]}}{\ProbQ[\ypzc_{t}]} \label{eq:DFR}.
\end{align}
An integral FR immediately follows as
\begin{align}
\average{e^{-\Delta \CMF[\ypzc_{t}]}}_{\ProbQ[\ypzc_{t}]}=1,\label{eq:IFR}
\end{align}
Furthermore,  we also have the Kawai-Parrondo-Broeck(KPB)-type FR as
\begin{align}
\average{\Delta \CMF_{t}} &
 =\KL{\hat{\ProbP}_{F}}{\ProbP_{F}} -\average{\KL{\hat{\ProbP}_{B}^{\ypzc}}{\ProbP_{B}^{\ypzc}}}_{\ProbQ} = \KL{\ProbQ}{\average{\hat{\ProbP}_{B}^{\xpzc}}_{\ProbP_{F}}} \label{eq:KBP}.
\end{align}
The second term shows that the average loss of a suboptimal strategy, $\average{\Delta \CMF}$, is determined by the strength of contraction of the phenotypic path probabilities from $\KL{\hat{\ProbP}_{F}}{\ProbP_{F}}$ to $\average{\KL{\hat{\ProbP}_{B}^{\ypzc}}{\ProbP_{B}^{\ypzc}}}_{\ProbQ[\ypzc]} $ that is induced by selection. 
The third term, in addition, shows that the loss is zero when $\average{\hat{\ProbP}_{B}^{\xpzc}[\ypzc_{t}]}_{\ProbP_{F}}$ equals to the statistics of environment, $\ProbQ[\ypzc_{t}]$.
In addition, this FRs can be used to quantify the loss by causal strategy even when $\hat{\ProbP}_{F}$ is not causal\cite{SM:2014}.
 
\submit{\section{Optimal Strategy with Environmental Signal}}
All the result above, i.e., the consistency condition, the maximal cumulative fitness, and the FRs, can be generalized for the situation where the environmental signal, $z_{t}$, is available.
Let us define the forward path probability with the signal as $\ProbP_{F}^{\zpzc}[\xpzc_{t}] \defeq \prod_{\tau=0}^{t-1}\ProbT(x_{\tau+1}|x_{\tau},z_{\tau+1})\ProbP_{F}(x_{0})$ for $\ProbT(x'|x,z)$. 
With this forward path probability, we similarly have the cumulative fitness with the signal, $\CMF[\ypzc_{t},\zpzc_{t}]= \ln\average{e^{\LF[\xpzc_{t},\ypzc_{t}]}}_{\ProbP_{F}^{\zpzc}[\xpzc_{t}]}$, and the backward path probability, $\ProbP_{B}^{\ypzc,\zpzc}[\xpzc_{t}] = e^{\LF[\xpzc_{t},\ypzc_{t}]-\CMF[\ypzc_{t}\zpzc_{t}]}\ProbP_{F}^{\zpzc}[\xpzc_{t}]$.
As in the case without the signal, we consider general $\ProbP_{F}^{\zpzc}$ than Markov or causal ones.
The joint and marginal backward probabilities are also defined as 
$\ProbP_{B}^{J}[\xpzc_{t},\ypzc_{t},\zpzc_{t}] \defeq \ProbP_{B}^{\ypzc,\zpzc}[\xpzc_{t}]\ProbQ[\ypzc_{t},\zpzc_{t}]$,
$\ProbP_{B}^{M}[\xpzc_{t},\zpzc_{t}]\defeq \sum_{\ypzc_{t}} \ProbP_{B}^{J}[\xpzc_{t},\ypzc_{t},\zpzc_{t}]$, 
$\ProbP_{B}^{M}[\xpzc_{t},\ypzc_{t}]\defeq \sum_{\zpzc_{t}} \ProbP_{B}^{J}[\xpzc_{t},\ypzc_{t},\zpzc_{t}]$, and $\ProbP_{B}^{M}[\xpzc_{t}]\defeq \sum_{\ypzc_{t},\zpzc_{t}} \ProbP_{B}^{J}[\xpzc_{t},\ypzc_{t},\zpzc_{t}]$.
Conditional probabilities are $\ProbP_{B}^{\zpzc}[\xpzc_{t},\ypzc_{t}]\defeq \ProbP_{B}^{J}[\xpzc_{t},\ypzc_{t},\zpzc_{t}]/\ProbQ[\zpzc_{t}]$, $\ProbP_{B}^{\ypzc}[\xpzc_{t},\zpzc_{t}]\defeq \ProbP_{B}^{J}[\xpzc_{t},\ypzc_{t},\zpzc_{t}]/\ProbQ[\ypzc_{t}]$,  $\ProbP_{B}^{M,\zpzc}[\xpzc_{t}]\defeq \ProbP_{B}^{M}[\xpzc_{t},\zpzc_{t}]/\ProbQ[\zpzc_{t}]$, and $\ProbP_{B}^{M,\ypzc}[\xpzc_{t}]\defeq \ProbP_{B}^{M}[\xpzc_{t},\ypzc_{t}]/\ProbQ[\ypzc_{t}]$.
With these extensions, the average cumulative fitness, $\average{\CMF_{t}} \defeq \average{\CMF[\ypzc_{t},\zpzc_{t}]}_{\ProbQ[\ypzc_{t},\zpzc_{t}]}$, is 
\begin{align*}
\average{\CMF_{t}} &=\average{\LF[\xpzc_{t},\ypzc_{t}]}_{\ProbP^{J}_{B}}-\MI_{B}^{\xpzc,\ypzc|\zpzc} -\average{\KL{\ProbP_{B}^{M,\zpzc}[\xpzc_{t}]}{\ProbP_{F}^{\zpzc}[\xpzc_{t}]}}_{\ProbQ[\zpzc_{t}]},
\end{align*}
where $\MI_{B}^{\xpzc,\ypzc|\zpzc} \defeq \average{\KL{\ProbP_{B}^{\zpzc}[\xpzc_{t}, \ypzc_{t}]}{\ProbP_{B}^{M,\zpzc}[\xpzc_{t}]\ProbQ[\ypzc_{t}|\zpzc_{t}]}}_{\ProbQ[\zpzc_{t}]}$.

We also have the optimal strategy with the signal as $\breve{\ProbP}_{F}^{\zpzc} \defeq \arg {\max}_{\ProbP_{F}^{\zpzc}} \average{\CMF_{t}}$ where we use $\breve{\,}$ to indicate the optimal strategy with the signal to distinguish it from one without the signal.
The optimal strategy satisfies the following extended consistency condition as \cite{SM:2014}
\begin{align}
\breve{\ProbP}^{M,\zpzc}_{B}[\xpzc_{t}]={\sum}_{\ypzc_{t}} \breve{\ProbP}_{B}^{\ypzc,\zpzc}[\xpzc_{t}]\ProbQ[\ypzc_{t}|\zpzc_{t}]=\breve{\ProbP}_{F}^{\zpzc}[\xpzc_{t}],\label{eq:consistencys}
\end{align}
and the corresponding maximal cumulative fitness is
\begin{align}
\average{\breve{\CMF}_{t}} &=\average{\LF[\xpzc_{t},\ypzc_{t}]}_{\breve{\ProbP}^{J}_{B}}-\breve{\MI}_{B}^{\xpzc,\ypzc|\zpzc}.\label{eq:maxacCMFs}
\end{align}
Similarly to the case without the signal, we can interpret this relation as an information compression of $\ypzc_{t}$ to $\xpzc_{t}$ with side information $\zpzc_{t}$\cite{SM:2014}.

By using $\breve{\ProbP}_{B}^{\xpzc,\zpzc}[\ypzc_{t}] \defeq\breve{\ProbP}_{B}^{J}/\breve{\ProbP}_{B}^{M}[\xpzc_{t},\zpzc_{t}]= e^{\LF[\xpzc_{t},\ypzc_{t}]-\breve{\CMF}[\ypzc_{t},\zpzc_{t}]}\ProbQ[\ypzc_{t}|\zpzc_{t}]$ derived from the extended consistency condition, we similarly have the fitness loss, $\Delta \CMF[\ypzc_{t},\zpzc_{t}]\defeq \breve{\CMF}[\ypzc_{t},\zpzc_{t}]-\CMF[\ypzc_{t},\zpzc_{t}]$, by a suboptimal strategy with signal, $\ProbP_{F}^{\zpzc}$ as
\begin{align}
e^{-\Delta \CMF[\ypzc_{t},\zpzc_{t}]} 
= \average{\breve{\ProbP}_{B}^{\xpzc,\zpzc}[\ypzc_{t}]}_{\ProbP_{F}^{\zpzc}[\xpzc_{t}]}/\ProbQ[\ypzc_{t}|\zpzc_{t}], \label{eq:DFRs0}
\end{align}

When $\hat{\ProbP}_{F}[\xpzc_{t}]$ shares the same support with $\breve{\ProbP}_{F}^{\zpzc}[\xpzc_{t}]$ , by choosing the optimal strategy without signal as $\ProbP_{F}^{\zpzc}[\xpzc_{t}]=\hat{\ProbP}_{F}[\xpzc_{t}]$,the Sagawa-Ueda detailed FR \cite{Sagawa:2012wi} as 
\begin{align}
e^{-(\hat{\CMF}[\ypzc_{t}]+i[\ypzc_{t},\zpzc_{t}]-\breve{\CMF}[\ypzc_{t},\zpzc_{t}])} 
= \hat{\ProbP}_{B}^{\xpzc}[\ypzc_{t}]/\breve{\ProbP}_{B}^{\xpzc,\zpzc}[\ypzc_{t}], \label{eq:DFRs1}
\end{align}
where $e^{i[\ypzc_{t},\zpzc_{t}]} \defeq \ProbQ[\ypzc_{t},\zpzc_{t}]/\ProbQ[\ypzc_{t}]\ProbQ[\zpzc_{t}]$.
The KPB-type FR, $\average{\hat{\CMF}}+\MI^{\ypzc,\zpzc}-\average{\breve{\CMF}}=\KL{\breve{\ProbP}_{B}^{J}}{\hat{\ProbP}_{B}^{\ypzc} \breve{\ProbP}_{B}^{M}}\ge 0$, shows that $\MI^{\ypzc,\zpzc}$ is an upper bound of the average gain of fitness by sensing.
Nonetheless, $\MI^{\ypzc,\zpzc}$ does not always properly quantify the gain of fitness by sensing. 
For example, if all phenotypes have identical growth under two environmental states, $y$ and $y'$,  i.e., $h(x,y)=h(x,y')$ for all $x$, the information in the signal to distinguish $y$ and $y'$ has no contribution to fitness gain whereas 
 $\MI^{\ypzc,\zpzc}$ increases.
 The information relevant for fitness can be evaluated more tightly by the following FR as 
\begin{align}
e^{-(\hat{\CMF}[\ypzc_{t}]+i[\ypzc_{t},\zpzc_{t}]-\breve{i}_{B}^{\xpzc}[\ypzc_{t},\zpzc_{t}]-\breve{\CMF}[\ypzc_{t},\zpzc_{t}])} = \frac{\hat{\ProbP}_{B}^{\xpzc}[\ypzc_{t}]}{\breve{\ProbP}_{B}^{M,\xpzc}[\ypzc_{t}]},\label{eq:DFRs2}
\end{align}
where $e^{\breve{i}_{B}^{\xpzc}[\ypzc_{t},\zpzc_{t}]} \defeq \breve{\ProbP}_{B}^{\xpzc,\zpzc}[\ypzc_{t}]/\breve{\ProbP}_{B}^{M,\xpzc}[\ypzc_{t}]$.
The KPB-type FR shows that the multivariate mutual information, $\breve{\MI}_{B}^{\xpzc,\ypzc,\zpzc}\defeq \MI^{\ypzc,\zpzc}-\breve{\MI}_{B}^{\ypzc,\zpzc|\xpzc}$, is the tighter bound for fitness gain by sensing as 
\begin{align}
\average{\hat{\CMF}}+\MI_{B}^{\xpzc,\ypzc,\zpzc}-\average{\breve{\CMF}}=\average{\KL{\breve{\ProbP}_{B}^{M,\xpzc}}{\hat{\ProbP}_{B}^{\xpzc}}}_{\breve{\ProbP}^{M}_{F}}\ge 0,\label{eq:KBPs2}
\end{align}
\cite{SM:2014}. 
In addition, the equality can be attained when the backward path probabilities of the optimal switching with and without sensing are identical as $\breve{\ProbP}_{B}^{M,\xpzc}=\hat{\ProbP}_{B}^{\xpzc}$.
Because $\breve{\MI}_{B}^{\ypzc,\zpzc|\xpzc}$ is the residual information of the signal on the environment when we already know the phenotype path (\fgref{fig2} (B)), $\breve{\MI}_{B}^{\xpzc,\ypzc,\zpzc}$ is the maximum information of the signal that can be imprinted into the phenotypic dynamics by selection, i.e., the information of the signal consumed and used in selection.

\submit{\section{Discussion}}
In this work, we derived various FRs for fitness loss and gain with and without sensing the environment.
These results generalize the previous results  obtained  by Kelly\cite{Rivoire:2011fy}, Hacco and Iwasa\cite{Haccou:1995tf}, and others for the average of fitness gain and loss.
In addition, by combining the FRs, we can also recover the result on the fitness gain by the optimal causal strategy derived in \cite{Rivoire:2011fy}(see \cite{SM:2014}).
The keystone for generalization was the introduction of path-wise formulation and the retrospective view of phenotypic dynamics via the backward path probability.
This also enables us to clarify an information-theoretic aspect of selection as passive compression of environmental dynamics onto the retrospective phenotypic one.
Active information processing by sensing interacts with this passive processing, and thereby, the maximum gain of fitness by sensing is quantified by the multivariate mutual information (\eqnref{eq:KBPs2}). 
Because of the shared mathematical structures, this work will be the basis for the integration of the information thermodynamics and evolutionary dynamics to unveil the interdependencies among fitness, information and entropy production\cite{Iwasa:1988ux,deVladar:2011kz,Qian:2014ha,Seifert:2012es, Sagawa:2012wi,Mustonen:2010ig}.

We thank Yoichi Wakamoto and Mikihiro Hashimoto for discussion.
This research is supported partially by Platform for Dynamic Approaches to Living System from MEXT, Japan, the Aihara Innovative Mathematical Modelling Project, JSPS through the FIRST Program, CSTP, Japan,  and the JST PRESTO program.




\end{document}